\documentstyle
[12pt]
{article}
\begin{document}

\newcommand{\bean}{\begin{eqnarray*}}
\newcommand{\eean}{\end{eqnarray*}}
\newcommand{\ed}{\end{document}}
\newcommand{\pr}{\prime}
\newcommand{\ppr}{\prime\prime}
\newcommand{\cE}{{\cal E}}
\newcommand{\vphi}{{\varphi}}
\newcommand{\oO}{O(k^{-1})}
\newcommand{\be}{\begin{equation}}
\newcommand{\ee}{\end{equation}}
\newcommand{\barr}{\begin{array}}
\newcommand{\earr}{\end{array}}
\newcommand{\bea}{\begin{eqnarray}}
\newcommand{\eea}{\end{eqnarray}}
\newcommand{\pa}{\partial}
\newcommand{\xx}{\hbox{}^*_*}

\title{Factorization of correlation functions in coset
conformal field theories.}

\author{A.V.Bratchikov  \thanks{bratchikov@kubstu.ru}  \\
 Kuban State Technological University,\\ 2 Moskovskaya Street,
     Krasnodar, 350072, Russia} \date {February,2000}

\maketitle

\begin{abstract}
We use the conformal Ward identities to study the structure
of correlation functions in coset conformal field theories.
For a large class of primary fields of arbitrary g/h theory a
factorization anzatz is found.Corresponding correlation functions
are explicitly expressed in terms of correlation functions of two
independent WZNW theories for g and h.
\end{abstract}



 Coset theories is an important subclass of two-dimensional
conformal invariant QFT's.(For a review see e.g.\cite {HKOC}.)
The $g/h$ coset theory is based on the Virasoro algebra
generated by \cite {
BH,GKO}
\bea \label {g/h}
K(m)=L^g(m)-L^h(m),\qquad  m\in Z.
\eea
The operator $L^g(m)$ is a conformal generator of the
Wess-Zumino-Novikov-Witten (WZNW)  theory [4-6] for
the Lie algebra $g$ and $h\subset g.$
In this paper we study the connection between the coset
and WZNW theories which follows from (\ref {g/h})
at the level of correlation functions and primary fields.

In the case of $su(2)/u(1)$ correlation functions
of primary fields may be written in terms of correlation functions
of the independent WZNW theories for $su(2)$ and $u(1)$ \cite
{FZ}.Correlation functions of the $g/u(1)^d, d=1\ldots rank\,g,$ cosets
\cite {B1} have a similar structure.In \cite {KLS}
some correlation functions of minimal models were expressed in terms of
correlation functions of two WZNW theories.

In this paper we show that a large class of correlation functions of
arbitrary $g/h$ coset conformal field theory can be expressed in terms
of correlation functions of two independent WZNW theories for $g$
and $h.$ To find correlation functions of coset primary fields we use
the conformal Ward identities \cite {BPZ}.We propose an anzatz for
coset primary fields and show that the corresponding correlation
functions satisfy the Ward identities. Different factorization
properties of $g/h$ coset correlation functions were found in \cite
{HO}.

The results of this paper are in agreement with those of refs.
\cite {GK,KPSY}  where the $g/h$ WZNW model was studied by using
path integral approach.

We begin with the affine Lie algebra $\hat g_k$ for simple $g$
\bean\label{alg}
[J^a(m),J^b(n)]=if^{abc}J^c{(m+n)}+{k}m\delta{^{ab}}\delta_{m+n,0},
\eean
where $f^{abc}$ are the structure constants of $g$ and $k$ is the
central charge.

The conformal generator $L^g(m)$ is given
by
\bean L^g(m)=\frac 1
{2k+Q_g}\sum_n{:J^a({m-n})J^a(n):},
\eean
where $Q_g$
is the quadratic Casimir in the adjoint representation of $g$.These
operators satisfy the commutator relations \bea \label {vir}
[L^g(m),L^g(n)]=(m-n)L^g(m+n)+c^g\left[\frac 1 {12}
(m^3-m)\right]\delta_{m,-n},
\eea
\bean
c^g=\frac {2k\,dim\,g} {2k+Q_g},
\eean
where $c^g$ is the central charge.

Let  $G_R(z)$ be the primary field of $\hat g_k$
\bea
[J^a(m),G_R(z)]=z^{m}G_R(z)t^a_R,
\label{al}
\eea
\bean
 [t^a_R,t^b_R]=if^{abc}t^c_R,
\eean
where $t^a_R$ is the representation of the generators
of $g$ for the field $G_R(z).$
In the WZNW theory $G_R(z)$ also is the primary field of
the Virasoro algebra (\ref {vir})
\bean \label{}
[L^g(m),G_R(z)]
=z^{m+1}\partial_z G_R(z) +\Delta_R (m+1)z^m
G_J(z),
\eean
\bean
\Delta_R=\frac {Q_J} {2k+Q_g},
\eean
where ${Q_R}$ is the quadratic Casimir of $g$
in the representation $R.$
Here and in what follows we treat only the holomorphic part.

Let $\hat h_k$ be a subalgebra of $\hat g_k.$ We assume that it is
generated by $J^A(m),A=1\ldots dim\,h.$
The field $G_R$ can be decomposed in
the set of some irreducible representations of $h$
\bea \label {dec}
G_R(z)=\sum_l{G_{R\,l}(z)}=\sum_l{P_lG_{R}(z)},
\eea
where $G_{R\,l}(z)$ belongs to the $l'$s representation and $P_l$ is the
corresponding projector.
The field $G_{R\,l}$ satisfies the equation
\bea \label {com1}
[J^A(m),G_{R\,l}(z)]=z^{m}G_{R\,l}(z)t^A_l,
\eea where $t^A_l$ is the representation of the
generators of $h$ for the field $G_{R\,l}(z)$
As well as  $G_{R}(z)$ the field $G_{R\,l}(z)$ is the primary field of
the Virasoro algebra (\ref{vir})
\bea
\label{L} [L^g(m),G_{R\,l}(z)] =z^{m+1}\partial_z G_{R\,l}(z)
+\Delta_R (m+1)z^m G_{R\,l}(z),
\eea
Correlation functions of these fields can be computed using
correlation functions of the WZNW theory
\bean \label{cb}
< G_{R_1l_1}(z_1)\ldots G_{R_Nl_N}(z_N)>=
\prod_{i=1}^N {P_{l_i}} < G_{R_1}(z_1)\ldots G_{R_N}(z_N)>
\eean

The coset conformal generators $K(m)$ (\ref {g/h}) satisfy
commutator relations (\ref {vir}) with the central
charge $c^{g/h}=c^g-c^h$ \cite {GKO}.
We shall need the relation
\newpage
\bean [K(m),G_{R\,l}(z)]=
z^{m+1} \left(\partial_z G_{R\,l}(z)- \frac 2  {2k+Q_h}
:J^A(z)G_{R\,l}(z):t^A_l\right)
\eean
\bea \label {ue}
+\Delta_{R\,l}(m+1)z^m G_{R\,l}(z)
,\;\;\;\;\;\;\;\;\;\;\;\;\;\;\
\eea
where
\bean :J^A(z)G_{R\,l}(z):=
\sum_{m<0}{ J^A(m) z^{-m-1}G_{R\,l}(z) +G_{R\,l}(z)\sum_{m\ge 0} J^A(m)
 z^{-m-1} }.
\eean
$\Delta_{R\,l}$ is given by
\bea\label{andim}
\Delta_{R\,l}=\Delta_R-\frac {Q_l} {2k+Q_h}, \eea
where ${Q_l}$ is the quadratic Casimir of $h$ in the representation $l.$

Correlation functions of the coset
primary  fields $\phi_i$ satisfy the conformal Ward identity \cite
{BPZ} \bea  < K(z)\phi_1(z_1)\ldots
\phi_N(z_N)>=
\sum_{i=1}^N
{
\left \{
\frac {\Delta_i} {(z-z_i)^2} +
\frac 1 {z-z_i} \frac {\pa} {\pa z{_i}}
\right\}
 <\phi_1(z_1)\ldots\phi_N(z_N)>},
 \label {WI}
\eea
where
$K(z)=\sum_m {K(m)z^{-m-2}}$ and
$\Delta_1,\Delta_2,\ldots,\Delta_N$ are dimensions of
$\phi_1,\phi_2,\ldots,\phi_N$, respectively.

To find a solution of this equation we shall use an
auxiliary WZNW theory.
Let $\hat h_{k'}$ be the auxiliary affine Lie algebra
\begin{equation}\label{alg}
[\chi^A(m),\chi^B(n)]=if^{ABC}\chi^C({m+n})+{k'}m
\delta{^{AB}}\delta_{m+n,0}.
\end{equation}
where $A,B,C=1\ldots dim\,h.$
The value of $ k' $ will be defined later.
Let $\Phi_l$ be the primary field of the WZNW theory for
$\hat h_{k'}$
\bea   \label {com2}
[\chi^A(m),\Phi_l(z)]=z^{m}\Phi_l(z)t^{*A}_l,
\eea
\bea   \label {KZ}
\frac {\pa} {\pa z}\Phi_l(z)= \frac {2}{2
k'+Q_h}:\chi^A(z)\Phi_l(z):t^{*A}_l,
 \eea where $t^{*A}_l=-(t^{A}_l)^T $ and $Q_h$ is the
quadratic Casimir in the adjoint
representation of $h.$ Eq.(\ref {KZ}) was introduced in ref. \cite
{KZ}.

We look for a solution of eq.(\ref {WI}) in the following factorized
form
\newpage
\bean \sum_{\alpha_1\ldots
 \alpha_N}<G_{R_1l_1}^{\alpha_1}(z_1)\ldots G_{R_Nl_N}^{\alpha_N}(z_N)
> <\Phi_{l_1}^{\alpha_1}(z_1)
\ldots \Phi_{l_N}^{\alpha_N}(z_N)>
\eean
\bea              \label {corr}
\equiv <\bigl(G{_{R_1l_1}}(z_1),\Phi{_{l_1}}(z_1)\bigr)
\ldots  \left(G{_{R_Nl_N}}(z_N),
\Phi{_{l_N}}(z_N)\right)>,
\eea
where $(\cdot,\cdot)$ is the  bilinear form
\bean
\left(G_{R\,l}(z),\Phi_{l}(z)\right)=
\sum_{\alpha=1}^{dim\,l}G_{R\,l}^{\alpha}(z)\Phi_{l}^{\alpha}(z).
\eean
We shall denote
\bea \label {tg}
\tilde G_{R\,l}(z)=\left(G{_{R\,l}}(z),\Phi{_{l}}(z)\right).
\eea
It follows from  (\ref {com1}) and (\ref{com2}) that
 $\tilde G_{R\,l}(z)$ commutes with the operator
$\tilde J^A(m)=J^A(m)+\chi^A(m)$
\bea \label {co}
[\tilde J^A(m),\tilde G_{R\,l}(z)]=0.
\eea

The vacuum state $\vert 0>$ is the joint state of the $\hat g_k$ and
$\hat h_{k'}$ WZNW theories.  We shall use the following properties of
$\vert 0>$ \bea \label {vac1} <0\vert \tilde J^A(m<0)= \tilde J^A(m \ge
0) \vert 0>=0, \eea \bea
\label {vac2} <0\vert K_-(z)= K_+(z)\vert 0> =0,  \eea where \bean
\label{} K_{-}(z)=\sum_{m<-1}K{(m)}z^{-m-2},\qquad K_{+}(z)=\sum_{m\ge
-1}K{(m)}z^{-m-2}.  \eean

Let us compute the left-hand side of eq.(\ref {WI}) using
correlation functions (\ref {corr}).
To simplify presentation we assume that $\vert z\vert >\vert
z_1\vert >\ldots > \vert z_N \vert $.  Using eqs.(\ref
 {vac2}), (\ref {ue}) and (\ref {KZ}) we get
\newpage
\bean   <K(z)\tilde
G_{R_1l_1}(z_1)\ldots \tilde G_{R_Nl_N}(z_N)>= <K_+(z)\tilde
G_{R_1l_1}(z_1)\ldots \tilde G_{R_Nl_N}(z_N)> \eean \bean =
\left\{\frac {\Delta_{R_1l_1}} {(z-z_1)^2}+ \frac {1} {z-z_1} \frac
{\pa} {\pa z{_1}}\right\} <\tilde G_{R_1l_1}(z_1)\ldots \tilde
 G_{R_Nl_N}(z_N)>\;\;\;\;\;\;\;\;\;\;\;\;\;\;\;\;\;\;\;\eean \bea +
 <\tilde G_{R_1l_1}(z_1)K_+(z)\ldots \tilde G_{R_Nl_N}(z_N) >+\frac {1} {z-z_1}
<T_{R_1l_1}(z_1)\ldots \tilde G_{R_Nl_N}(z_N) >,\label{step}
\eea
where
\bean
T_{R_1l_1}(z)=\frac {1} {2k'+Q_h}
(G_{R_1l_1}(z)t^A,:\chi^A(z)\Phi_{l_1}(z):)
-\frac {1}
{2k+Q_h} (:J^A(z)  G_{R_1l_1}(z):t^A,\Phi_{l_1}(z)).\label
{T}
\eean

At  $k'=-k- Q_h$ the field $T_{R_1l_1}(z)$
can be
written in the form \bean T_{R_1l_1}(z)=-\frac {1}{2k+Q_h} {:\tilde
J^A(z)(G_{R_1l_1}(z)t^A_{l_1},\Phi_{l_1}(z)):}.
\eean
Due to eqs. (\ref {co}) and (\ref {vac1}) the last term of eq.(\ref
{step}) vanishes
\bean
<T_{R_1l_1}(z_1)\ldots \tilde G_{R_Nl_N}(z_N) >=0.
\eean
Proceeding inductively one can show that the
correlation function (\ref {corr}) satisfies the Ward identity (\ref
{WI}).

From the arguments presented above it follows that  $\tilde
G_{R\,l}(z)$ (\ref {tg}) represents the primary field of the $g/h$
coset theory which has the conformal dimension (\ref {andim}).We took
the fields $G_{R\,l}(z)$ from the decomposition (\ref {dec}). However
to prove the factorization only eqs. (\ref {com1}) and  (\ref {L}) were
essentially used.  These equations have  other solutions which can be
used to construct coset primary fields.

To construct coset currents let us consider the field $J(z)= (J^i(z)),
J^i(z)=\sum_mJ^i_m z^{-m-1},i=dim\,h+1 \ldots  dim\, g.$ It can be
decomposed in the set of some irreducible representations of $h$ \bean
\label {e} J(z)=\sum_s{{J_s}(z)}.  \eean The field ${J_s}(z)$ satisfies
eq.(\ref {com1}) for some $t^A_s$ and eq.(\ref {L}) with the conformal
dimension  $\Delta=1.$ According to eq. (\ref {tg}) the coset current
corresponding to ${J_s}(z)$ is given by \bea \label {current} \tilde
J_s(z)= (J_s(z),\Phi_s(z)).  \eea It follows from (\ref {andim}) that
$\tilde J_s(z)$ has the conformal dimension \bean  1- \frac {Q_s}
{2k+Q_h}, \eean where ${Q_s}$ is the quadratic Casimir of $h$ in the
representation $s.$

Let us consider the $g/u(1)^d,1\le d\le rank\,g,$ coset
theory.In this case the primary field $G_R(z)$ is decomposed in the set
of one-dimensional representations of $u(1)^d$ \bean \label {}
G_R(z)=\sum_{\mu=1}^{dim\,R}{G_{R\mu}(z)},
\eean
\bean
[J^A(m),{G_{R\mu}(z)}]= \mu^Az^m{G_{R\mu}(z)},
\eean where
$\mu=(\mu^A).$
      A solution of eqs.(\ref {com2}),( \ref {KZ}) is given by
\bea
\Phi_\mu(z)= :exp\left(-{i\over {k'}}\mu\cdot\varphi(z)\right):,
\eea
\begin{equation}\label{boson}
\varphi^{A}(z)=q^{A}-i\chi^A(0)logz+i\sum_{n\not
=0}{{\chi^A(n)}\over n}z^{-n},
\end{equation}
where
\begin{equation}\label{gamma}
[q^{A},\chi^B({m})]=i\delta^{AB}\delta_{m,0}.
\end{equation}

According to eqs.(\ref {tg}) and (\ref {andim}) at $k'=-k$ $\tilde
G_{R\mu}(z)= G_{R\mu}(z)\Phi_\mu(z)$ represents the coset primary field
which has the dimension  $\Delta_{R\mu}=\Delta_R-{\mu^2}/ {2k}.$

The correlation function of these fields is given by
\bean \label{}
< G_{R_1\mu_1}(z_1)\ldots G_{R_N\mu_N}(z_N)>=
< G_{R_1}(z_1)\ldots G_{R_N}(z_N)>\prod_{i<j}^N(z_i-z_j)^{-\frac
{\mu_i\mu_j} k}
. \eean This is in agreement with the results of
refs.\cite {FZ,B1}. Parafermion $g/u(1)^d$  currents in the form (\ref
{current}) were obtained in \cite {B2}.

The results presented in this paper can be extended in many
directions. The most important is to study the factorization
properties of  the  $W/h$ coset conformal field theory.
It is also interesting to find primary fields
and describe the corresponding operator algebra. This is presently
being studied.

\end{document}